\documentclass[twocolumn]{aastex62}
\usepackage{graphicx}
\usepackage{natbib}
\usepackage{placeins}
\usepackage{amsmath}
\usepackage{tgcursor}
\usepackage{booktabs}
\usepackage[caption=false]{subfig}

\setcitestyle{notesep={ }}

\begin{document}

\title{Detection of phosphorus-bearing molecules towards a Solar-type protostar}

\author{Jennifer B. Bergner}
\affiliation{Harvard-Smithsonian Center for Astrophysics, 60 Garden Street, Cambridge, MA 02138, USA}

\author{Karin I. \"Oberg}
\affiliation{Harvard-Smithsonian Center for Astrophysics, 60 Garden Street, Cambridge, MA 02138, USA}

\author{Salma Walker}
\affiliation{California State University, Northridge, CA 91330, USA}

\author{Viviana V. Guzm\'an}
\affiliation{Instituto de Astrof\'isica, Pontificia Universidad Cat\'olica de Chile, Av. Vicu\~na Mackenna 4860, 7820436 Macul, Santiago, Chile}

\author{Thomas S. Rice}
\affiliation{Department of Astronomy, Columbia University, 550 W 120th Street, New York, NY 10027, USA}

\author{Edwin A. Bergin}
\affiliation{Department of Astronomy, University of Michigan, 500 Church Street Ann Arbor, MI 48109, USA}

\begin{abstract}
\noindent Phosphorus is a key ingredient in terrestrial biochemistry, but is rarely observed in the molecular ISM and therefore little is known about how it is inherited during the star and planet formation sequence.  We present observations of the phosphorus-bearing molecules PO and PN towards the Class I low-mass protostar B1-a using the IRAM 30m telescope, representing the second detection of phosphorus carriers in a Solar-type star forming region.  The P/H abundance contained in PO and PN is $\sim$10$^{-10}$--10$^{-9}$ depending on the assumed source size, accounting for just 0.05--0.5\% of the solar phosphorus abundance and implying significant sequestration of phosphorus in refractory material.  Based on a comparison of the PO and PN line profiles with the shock tracers SiO, SO$_2$, and CH$_3$OH, the phosphorus molecule emission seems to originate from shocked gas and is likely associated with a protostellar outflow.  We find a PO/PN column density ratio of $\sim$1--3, which is consistent with the values measured in the shocked outflow of the low-mass protostar L1157, the massive star-forming regions W51 and W3(OH), and the galactic center GMC G+0.693-0.03.  This narrow range of PO/PN ratios across sources with a range of environmental conditions is surprising, and likely encodes information on how phosphorus carriers are stored in grain mantles.
\end{abstract}


\section{Introduction}

Phosphorus is a critical bioelement on Earth, and is one of just a few elements that appears ubiquitously in all known life forms.  Despite being relatively scarce on a cosmic scale \citep[P/H $\sim$2.6$\times$10$^{-7}$;][]{Asplund2009} , the unique bonding properties of phosphorus have made it a key component of various biochemical functionalities, from membrane formation to energy storage to genetic encoding  \citep{Macia2005}.  Given the special role of phosphorus in terrestrial biochemistry, the availability of phosphorus on other planets may be a precondition for whether origins of life chemistry is possible.

At present, there are very few constraints on the phosphorus chemistry during low-mass star formation (and by extension planet formation).  Small P-bearing molecules including PO, PN, CP, and HCP have been detected towards evolved stars \citep{Agundez2007, Tenenbaum2007, Ziurys2007, Milam2008}, but attempts to detect these molecules towards active star-forming regions have had mixed results.  In massive star forming regions, PN has been found fairly commonly \citep{Turner1987, Fontani2016, Mininni2018}, while PO has been detected towards only two high-mass star forming regions despite numerous searches \citep{Matthews1987, Fontani2016, Rivilla2016}. In low-mass star forming regions, PN and PO were detected in the outflow shock of the protostar L1157, but not towards the protostellar envelope \citep{Yamaguchi2011, Lefloch2016}.  When detected around low- and high-mass protostars alike, phosphorus is found to be depleted from the volatile phase by one to two orders of magnitude \citep[e.g.][]{Turner1987, Rivilla2016, Lefloch2016}, suggesting substantial sequestration in the solid state. 

As a result of the scarcity of phosphorus molecule detections, particularly during low-mass star formation, the phosphorus chemistry in planetary system progenitors is poorly constrained.  Here we present detections of the phosphorus carriers PN and PO towards the Class I Solar-type protostar B1-a with the IRAM 30m telescope.  PO and PN lines were serendipitously detected in one source out of a sample of 16 embedded low-mass protostars \citep[the full sample is described in][]{Graninger2016}.  Combining these measurements with subsequent follow-up observations, we detect three PN lines and eight PO lines in B1-a, enabling a rotational diagram analysis to constrain the column densities and excitation temperatures of PO and PN.  We discuss implications for the phosphorus chemistry based on line profile analysis and the PO/PN ratio, and make comparisons with the Solar nebula based on meteoritic and cometary measurements.

\section{Observations}
\label{sec:obs}
Observations of B1-a (J2000 R.A. = 03:33:16.67, Decl. = 31:07:55.1) were taken with the IRAM 30m telescope.  The EMIR 90 GHz (3 mm), 150 GHz (2 mm), and 230 GHz (1 mm) receivers were used with the Fourier Transform Spectrometer (FTS) backend.  The 3 mm observations were taken 2013 July 17 with a 200 kHz resolution; the 2 mm observations were taken on 2018 March 26 with a 50 kHz resolution; and the 1 mm observations were taken on 2018 April 17-19 with a 200 kHz resolution.  The telescope half-power beam width is 27\arcsec, 16\arcsec, and 11\arcsec at 90 GHz, 150 GHz, and 230 GHz, respectively.  Initial data reduction was performed in CLASS\footnote{http://www.iram.fr/IRAMFR/GILDAS/}.  Spectra were then exported for subsequent analysis with Python. 

\section{Results}
\subsection{PO and PN detections}
\label{sec:res_dets}

\begin{figure}
\begin{centering}
	\includegraphics[width=\linewidth]{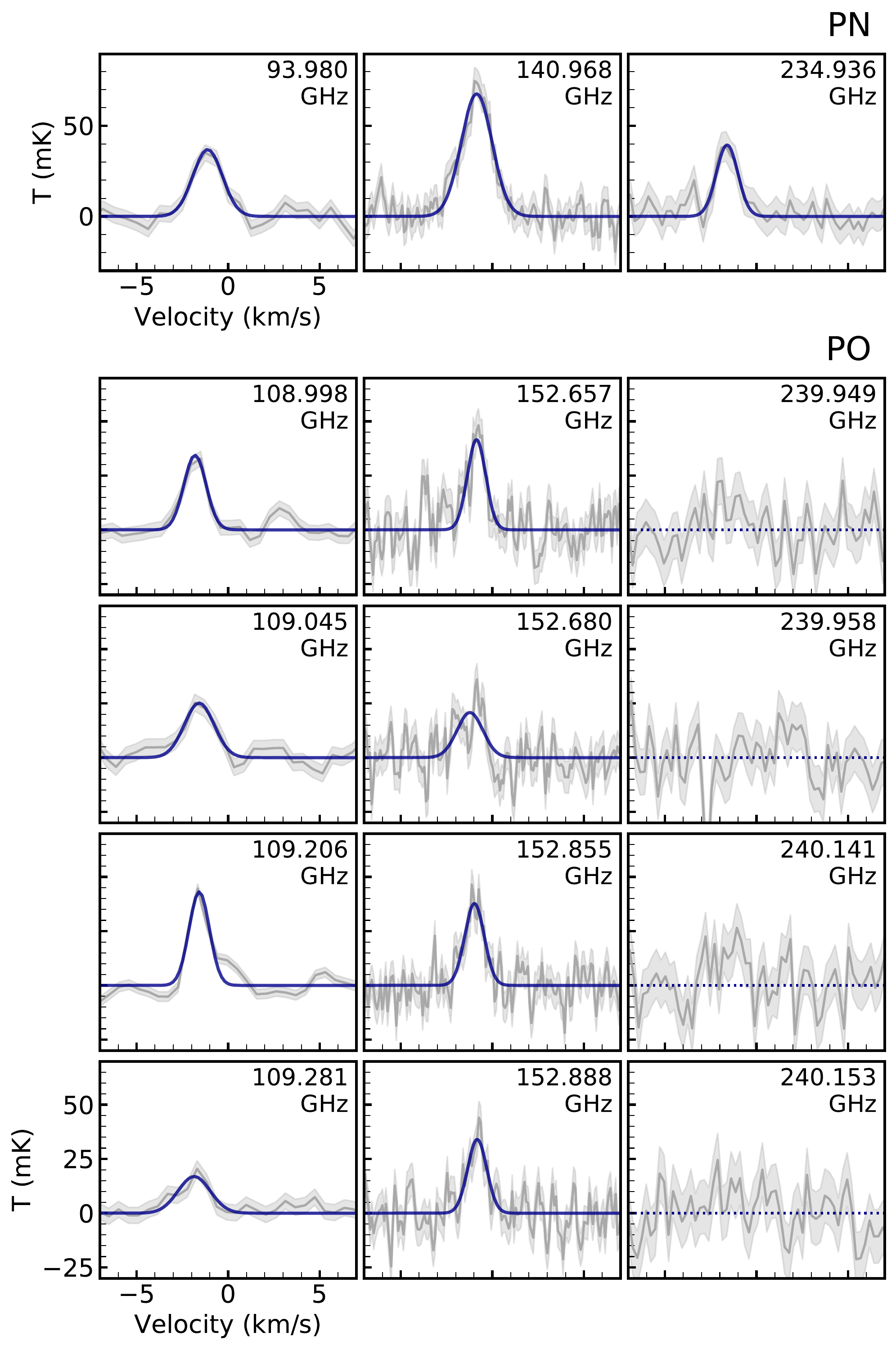}
	\caption{PN and PO transitions observed towards B1-a are shown as grey lines, with shaded regions representing the rms.  Gaussian fits to detected lines are shown in blue.}
	\end{centering} 
	\label{fig:spec_fits}
\end{figure}

In each of the 1 mm, 2 mm, and 3 mm setups, we cover a single PN line and four PO hyperfine components.  Spectral line parameters are taken from the CDMS catalog \citep{Muller2001, Muller2005} with data from \citet{Kawaguchi1983}, \citet{Bailleux2002}, and \citet{Cazzoli2006}, and are listed in Table \ref{tab:linedat}.  All three targeted PN lines are detected, as well as eight total PO lines in the 3 mm and 2 mm setups.  When the PO 6$_{-1,6,6}$--5$_{1,5,5}$ and 6$_{1,6,6}$--5$_{-1,5,5}$ spectra in the 1 mm setup are stacked, we obtain a marginal 2.7$\sigma$ detection ($\int T_{mb} dV$ = 30.8 $\pm$ 11.1 mK km s$^{-1}$), which is treated as an upper limit for all subsequent analysis.  The PN and PO line targets, along with Gaussian fits for detections, are shown in Figure \ref{fig:spec_fits}.  The velocity-integrated main-beam temperatures for each line are listed in Table \ref{tab:linedat}; for subsequent analysis, uncertainties consist of the Gaussian fit uncertainties added in quadrature with a 10\% calibration uncertainty.

\begin{deluxetable}{lcrrrrr} 
	\tabletypesize{\footnotesize}
	\tablecaption{Spectral line data$^a$ \label{tab:linedat}}
	\tablecolumns{7} 
	\tablewidth{\textwidth} 
	\tablehead{
		\colhead{}           &
		\colhead{Transition}         &
		\colhead{Freq.}       &
		\colhead{$E_u$}              &
		\colhead{S$\mu^2$}        &
		\colhead{$g_u$}              &
		\colhead{$\int T_{mb} dV$ $^b$}  \\
		\colhead{}                        & 
		\colhead{}                        & 
		\colhead{(GHz)}               &
		\colhead{(K)}                   &
		\colhead{(D$^2$)}            &
		\colhead{}                         & 
		\colhead{(mK km s$^{-1}$)}                                        
		}
\startdata
PN & 2 -- 1 & 93.980 & 6.8 & 15.1 &   5 & 76.5 [14.1]  \\ 
 & 3 -- 2 & 140.968 & 13.5 & 22.6 &   7 & 142.2 [16.2]  \\ 
 & 5 -- 4 & 234.936 & 33.8 & 37.7 &  11 & 58.9 [11.9]  \\ 
 \hline
PO & 3$_{1,3,3}$ -- 2$_{-1,2,2}$ & 108.998 & 8.4 & 9.9 &   7 & 49.3 [8.7]  \\ 
 & 3$_{1,3,2}$ -- 2$_{-1,2,1}$ & 109.045 & 8.4 & 6.4 &   5 & 47.6 [11.4]  \\ 
 & 3$_{-1,3,3}$ -- 2$_{1,2,2}$ & 109.206 & 8.4 & 9.9 &   7 & 59.8 [10.4]  \\ 
 & 3$_{-1,3,2}$ -- 2$_{1,2,1}$ & 109.281 & 8.4 & 6.4 &   5 & 38.4 [10.6]  \\ 
 & 4$_{-1,4,4}$ -- 3$_{1,3,3}$ & 152.657 & 15.7 & 13.6 &   9 & 52.3 [9.0]  \\ 
 & 4$_{-1,4,3}$ -- 3$_{1,3,2}$ & 152.680 & 15.7 & 10.1 &   7 & 36.5 [8.5]  \\ 
 & 4$_{1,4,4}$ -- 3$_{-1,3,3}$ & 152.855 & 15.8 & 13.6 &   9 & 51.3 [8.2]  \\ 
 & 4$_{1,4,3}$ -- 3$_{-1,3,2}$ & 152.888 & 15.7 & 10.1 &   7 & 45.1 [8.3]  \\ 
 & 6$_{-1,6,6}$ -- 5$_{1,5,5}$ & 239.949 & 36.7 & 20.9 &  13 & $<$52.6  \\ 
 & 6$_{-1,6,5}$ -- 5$_{1,5,4}$ & 239.958 & 36.7 & 17.4 &  11 & $<$31.3  \\ 
 & 6$_{1,6,6}$ -- 5$_{-1,5,5}$ & 240.141 & 36.7 & 20.9 &  13 & $<$53.1  \\ 
 & 6$_{1,6,5}$ -- 5$_{-1,5,4}$ & 240.153 & 36.7 & 17.4 &  11 & $<$69.5  \\ 
 \hline
 \hline
  SiO & 2--1 & 86.847 & 6.3 & 19.2 & 5 & - \\
  & 5--4 & 217.105 & 31.3 & 48.0 & 11 & 751 [94] \\
  & 6--5 & 260.518 & 43.8 & 57.6 & 13 & - \\
 SO$_2$ & 6$_{2.4}$--6$_{1,5}$ & 140.306 & 29.2 & 10.2 & 13 & 93 [12] \\
 CH$_3$OH & 2$_{-0,2}$--2$_{1,2}$ E & 157.276 & 20.1 & 9.6 & 20 & 745 [94] \\
\enddata
\tablenotetext{}{($a$) All line parameters are taken from the CDMS catalogue. ($b$) Uncertainties are listed in brackets.  Upper limits are 3$\sigma$.}
\end{deluxetable}

\subsection{Column densities and abundances}
\label{sec:res_cd}

Our observations cover multiple transitions of PO and PN, enabling a derivation of rotational temperatures and column densities using the rotational diagram method \citep{Goldsmith1999}.  Details on the rotational diagram fitting method can be found in Appendix \ref{sec:app_rd}.  The largest uncertainty in fitting the PO and PN rotational diagrams is that the degree of beam dilution is difficult to constrain from single-dish observations.  We therefore perform the rotational diagram analysis assuming a range of source sizes $R_{src}$ from 2\arcsec to 12\arcsec, corresponding to $\sim$600 --  3600 AU at the distance to Perseus \citep{Ortiz2018}.  This spans the range from compact emission around the central protostar to diffuse envelope emission (see Section \ref{sec:disc_structure} for a discussion of the source structure).  In each case, the beam dilution for an observed transition is calculated from:

\begin{equation}
\eta_{bf} = \frac{(R_{src})^2}{(\theta_b/2)^2 + (R_{src})^2},
\label{eq:bf}
\end{equation}

\noindent where $\theta_b$ is the beam FWHM in arcseconds (approximated as 2460 GHz/$\nu$ for the IRAM 30m telescope).  We note that this treatment assumes a Gaussian source shape, while the actual source structure may be more complex; a more sophisticated treatment requires resolved observations of the PO and PN emission.

\begin{figure}
\begin{centering}
	\includegraphics[width=0.85\linewidth]{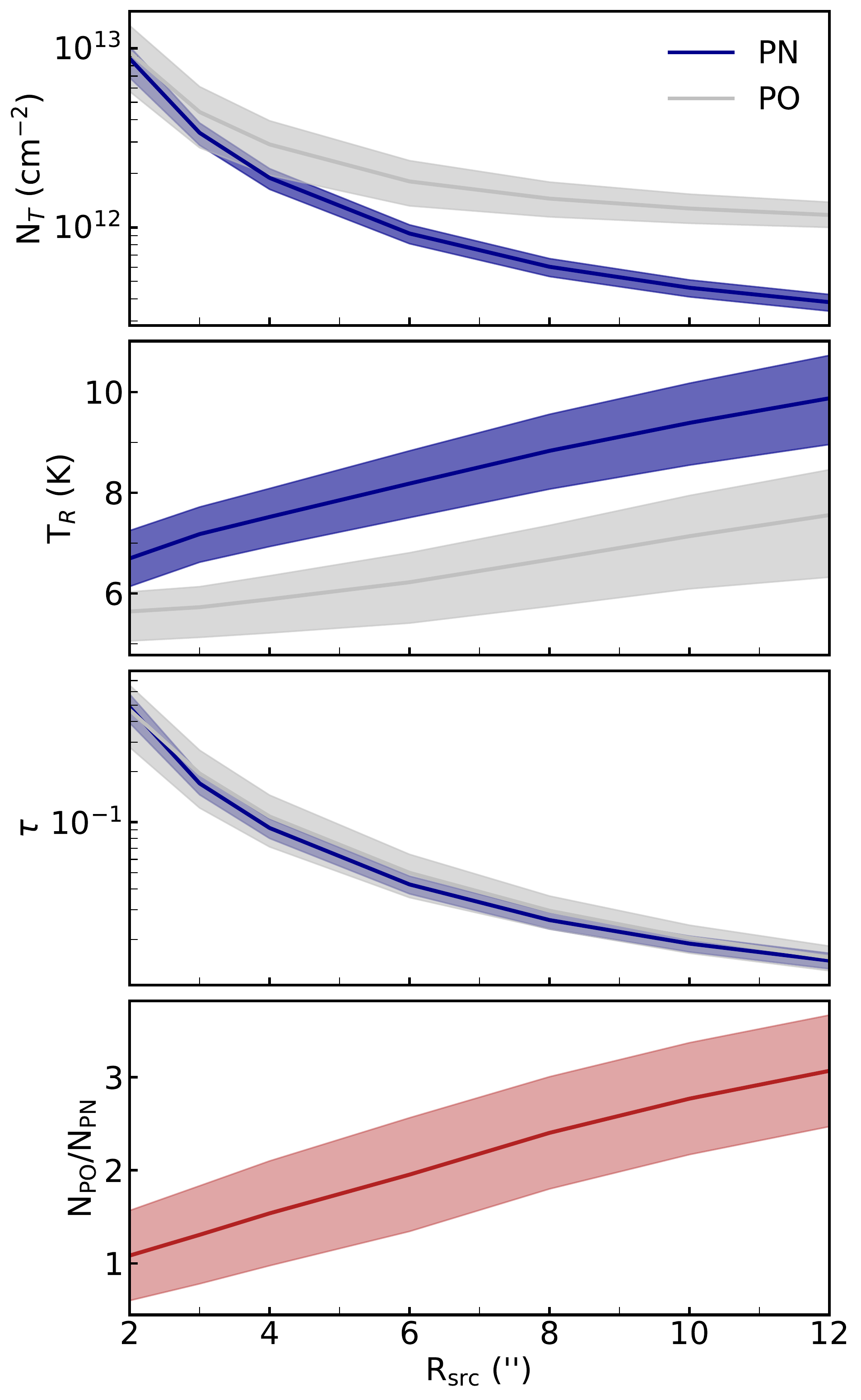}
	\caption{Rotational diagram results as a function of the assumed source size.  From top to bottom: PN and PO column densities, PN and PO rotational temperatures, optical depth of the optically thickest PO and PN line, and the PO/PN column density ratio.  Shaded regions show 1$\sigma$ uncertainties.}
	\end{centering} 
	\label{fig:profs}
\end{figure}

Figure \ref{fig:profs} shows the results of the rotational diagram analysis.  We find column densities between $\sim$10$^{11}$--10$^{13}$ cm$^{-2}$ for PN and 10$^{12}$--10$^{13}$ cm$^{-2}$ for PO, and rotational temperatures between $\sim$6 and 10 K for both molecules.  Importantly, for all assumptions of source size, the PO and PN lines with the highest optical depth remain optically thin.  The integrated intensity ratios of the PO hyperfine components are also consistent within the uncertainties with the intrinsic line strength ratios (Table \ref{tab:linedat}), again indicative of optically thin emission.  This means that optical depth effects are not an issue in deriving PO/PN ratios.  PO/PN ratios vary from $\sim$1--3 for different source size assumptions, but in all cases the ratio is $>$1.  

To derive the elemental abundance of phosphorus contained in the phosphorus carriers PO and PN, we require an estimate of the H$_2$ column density.  We use the hyperfine-split C$^{17}$O 1--0 transition to determine the CO column density, and in turn the H$_2$ column density, as described in detail in Appendix \ref{sec:app_CO}.  We find that the P/H abundance contained in PO and PN is 1.4$\times$10$^{-9}$ assuming the phosphorus molecules emit from a 2\arcsec region, or 1.2$\times$10$^{-10}$ assuming a 12\arcsec emission region.  We note this treatment assumes that the C$^{17}$O 1--0 emission uniformly fills the beam; if there is C$^{17}$O beam dilution in our observations then the true phosphorus abundances would be even lower.

\subsection{Non-LTE effects}
\label{sec:res_nonlte}
The critical densities of the targeted PN lines are high ($\sim$10$^5$--10$^{7}$ cm$^{3}$) compared to typical protostellar environments, and it is therefore likely that the emission is sub-thermal.  This is also consistent with the very low ($\sim$6--10 K) rotational temperatures derived for both molecules.  To evaluate the impact of non-LTE effects on our observations, we use the radiative transfer code \texttt{RADEX} \citep{vanDerTak2007} along with PN collisional rates taken from \citet{Tobola2007} via the BASECOL database \citep{Dubernet2013} to explore PN excitation in different density and temperature conditions.

We run a grid of \texttt{RADEX} models with gas kinetic temperatures ranging from 5--40 K and gas densities from 10$^4$--10$^8$ cm$^{-3}$, conditions typical of a low-mass protostellar environment.  We adopt a fixed line width of 1.5 km/s based on the observed line profiles, and a PN column density of 10$^{12}$ cm$^{-2}$.  For each gas temperature and gas density combination, we obtain a synthetic velocity-integrated intensity for the three observed PN lines.  We then perform rotational diagram analysis for the simulated observations to derive synthetic rotational temperatures and column densities.  

\begin{figure}
\begin{centering}
\vspace{0.1in}
	\includegraphics[width=\linewidth]{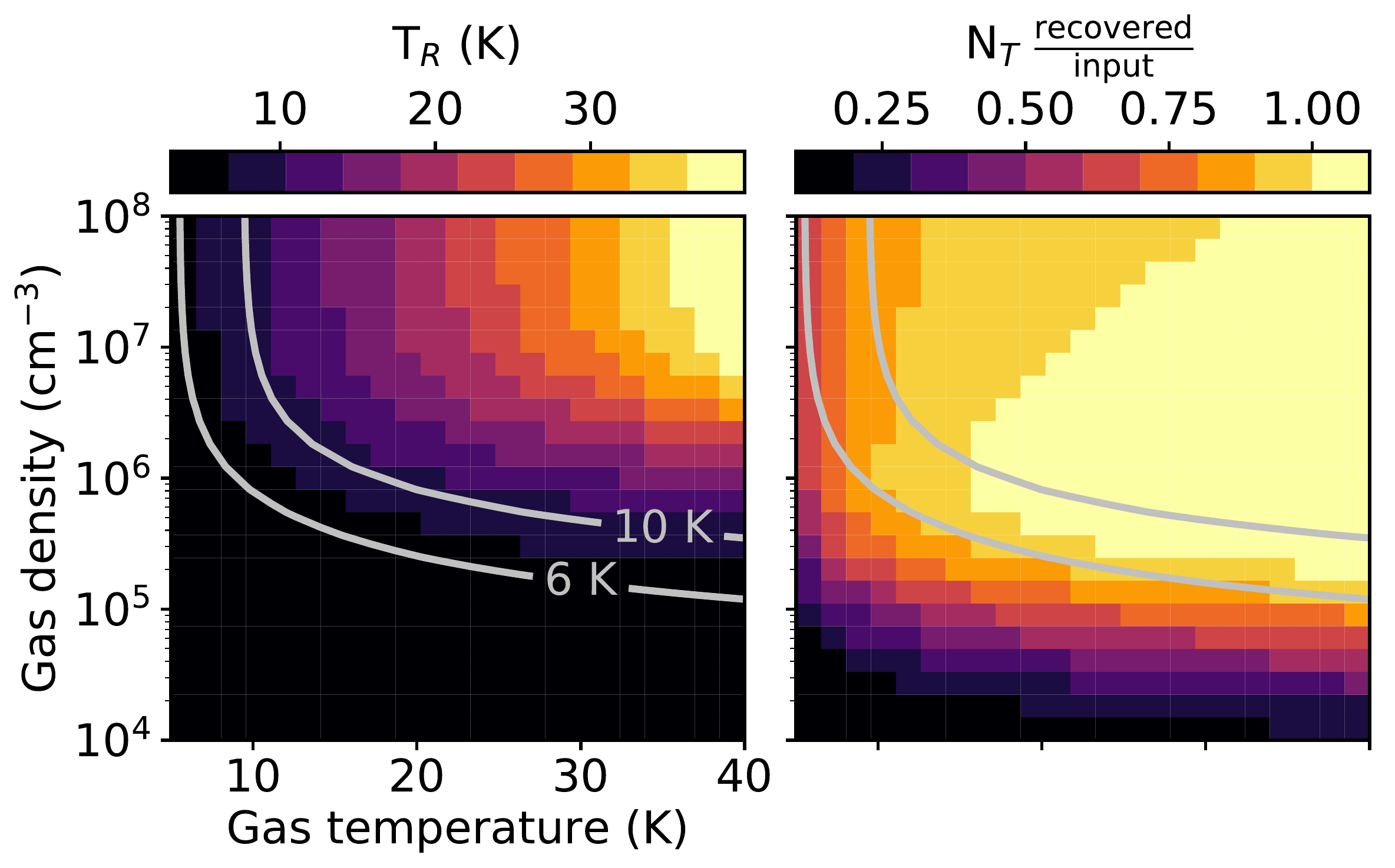}
	\caption{\texttt{RADEX} modeling results as a function of input gas density and temperature for an input PN column density of 10$^{12}$ cm$^{-2}$.  Left: recovered rotational temperatures, with contours marking the region consistent with the observed rotational temperatures of 6--10 K.  Right: recovered column density as a fraction of the input column density, with contours reproduced from the left panel.}
	\end{centering} 
	\label{fig:radex}
\end{figure}

Figure \ref{fig:radex} (left) shows the resulting rotational temperatures from this analysis, with the 6 and 10 K contours highlighting the region of parameter space consistent with our observationally derived rotational temperatures.  We recover thermal PN rotational temperatures only at gas densities above $\sim$10$^7$ cm$^{-3}$.  Assuming a power-law density profile for a typical low-mass protostar $n_\mathrm{H}(r)$ = 10$^6$ cm$^{-3}$ ($r$/1000 AU)$^{-3/2}$ \citep[adapted from][]{Jorgensen2002}, a density of 10$^7$ cm$^{-3}$ occurs only within the inner $\sim$215 AU of the central protostar.  Given a source luminosity of 1.3 L$_\odot$ \citep{Hatchell2007} and the temperature profile from \citet{Chandler2000}, the temperature should be above $\sim$35 K at these radii.  We can therefore rule out that the PN emission originates from the protostellar core (i.e., an environment that is both dense and warm); rather, it likely emits sub-thermally from lower-density gas, which may be either cool or warm. 

Figure \ref{fig:radex} (right) shows how the recovered PN column densities compare to the input column density, with the same parameter space contours reproduced from the left panel.  For the parameter space that is consistent with our derived rotational temperatures, we find that non-LTE effects on the derived column densities are quite low: the recovered column densities typically deviate $<$25\% from the input value.  We have also tested input column densities from 10$^{11}$ to 10$^{13}$ cm$^{-2}$, i.e. the range of observationally derived column densities, and find a similar level of agreement.  Thus, we expect that non-LTE effects have a small impact on our derived column densities compared to other sources of uncertainty and error.

\section{Discussion}

\subsection{Source structure \& emission origin}
\label{sec:disc_structure}

\begin{figure*}
\begin{centering}
	\includegraphics[width=\linewidth]{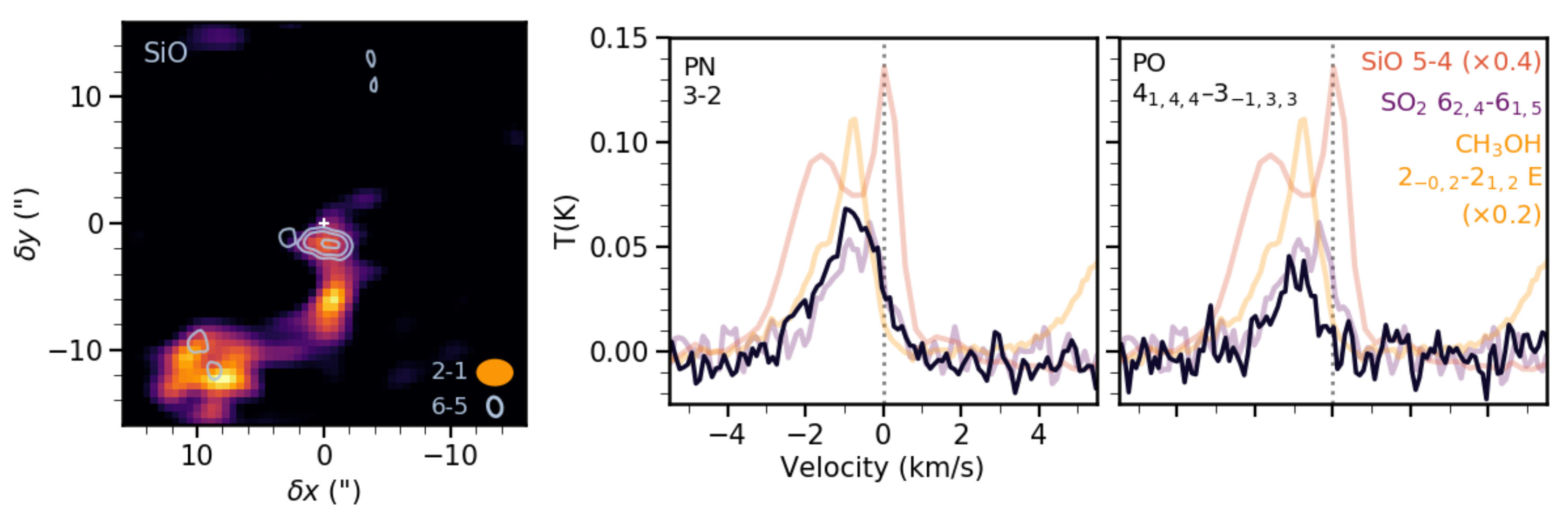}
	\caption{Left: Integrated intensity maps of SiO in B1-a taken with the NOEMA interferometer.  The continuum center is marked with a white `+'.  Colors show the 2--1 transition (integrated from -8.5--20.5 km/s) and contours show the 6--5 transition (integrated from -1.7--7.2 km/s) at 5, 10, and 20$\sigma$ levels, where $\sigma$ = 17.5 mJy/beam.  Restoring beams for each map are indicated in the bottom right.  Right: IRAM 30m spectra of the PN and PO transitions with the highest SNR and spectral resolution in the data set.  In each panel, black lines show the PN or PO transition.  For comparison, the SiO 5--4, SO$_2$ 6$_{2,4}$--6$_{1,5}$, and CH$_3$OH 2$_{-0,2}$--2$_{1,2}$ E transitions are shown in both panels in pink, purple, and yellow, respectively.  The SiO and CH$_3$OH lines are scaled for clarity.  Spectral parameters for each line can be found in Table \ref{tab:linedat}.}
	\end{centering} 
	\label{fig:kinematics}
\end{figure*}

Phosphorus molecules are often associated with shocks when detected in the dense ISM \citep[e.g.][]{Yamaguchi2011, Rivilla2018}, though recent studies of PN emission in massive dense cores indicate a more quiescent origin in some sources \citep{Rivilla2016, Mininni2018}.  Understanding whether the phosphorus molecule emission in B1-a is associated with quiescent gas (e.g. the envelope) or shocked gas (e.g. outflows) holds clues to how phosphorus molecules are released into the gas in low-mass star-forming environments.  While our PO and PN observations are spatially unresolved, we can use line kinematics to constrain their emission origin within B1-a.  Additionally, we have complementary images of the shock tracer SiO taken with the NOEMA interferometer that offer insight into the shocked outflow structures in the source.  Details on the NOEMA observations can be found in Appendix \ref{sec:app_noema}.

Figure \ref{fig:kinematics} (left) shows the integrated intensity maps of SiO 2--1 and 6--5 taken from the NOEMA observations.  Both lines trace an apparent outflow from the protostellar core.  The 2--1 line emission projects south of the source center, with an additional component extending to the southwest.  The 6--5 line shows compact emission just to the south of the continuum center, and weak emission in the southwest.  The difference in the emission regions of the two lines may simply be an excitation effect relating to the higher upper energy of the 6--5 line ($E_u$ = 44 K) compared to the 2--1 line ($E_u$ = 6 K).

In Figure \ref{fig:kinematics} (right), the spectral profile of the SiO 5--4 transition observed with the IRAM 30m telescope is shown.  There are two kinematic components: a strong feature at the source rest velocity, and a weaker, broader feature at blue-shifted velocities.  These features may correspond to the central and southwest emission components seen for the SiO 6--5 line (Figure \ref{fig:kinematics} left), though the spectral resolution of the NOEMA data is too low to confirm.  It is also possible that the depression in the SiO line profile is due to self-absorption in one continuous broad feature, rather than two distinct components.

The PO and PN lines with the highest spectral resolution and SNR in our data set are also shown in Figure \ref{fig:kinematics} (right).  Both lines are offset to blue-shifted velocities from the source rest velocity, coincident with the center of the SiO 5--4 line.  Our observations also cover lines of the weak-shock tracers CH$_3$OH and SO$_2$, which also share similar kinematics to PO and PN (Figure \ref{fig:kinematics} right). Thus, the phosphorus molecules appear to originate from shocked gas, likely within the outflow traced by SiO in Figure \ref{fig:kinematics} (left).  

The PN and PO line shapes are not well matched to the SiO 5--4 profile, which may be either due to a release of phosphorus molecules under different shock conditions (and therefore in different parts of the outflow), or due to excitation effects.  The latter is likely the case since the PN and PO lines shown in Figure \ref{fig:kinematics} have upper energies of 13 K and 16 K, respectively, while the SiO 5--4 line has an upper energy of 31 K, and should therefore be localized to a more compact emission region around the protostellar core.  We expect that the phosphorus molecule emission is spatially similar to the SiO 2--1 emission ($E_u$ = 6 K) shown in Figure \ref{fig:kinematics} (left), though resolved observations are needed to fully explore the spatial distribution of phosphorus molecules in this source.

\citet{Lefloch2016} also detected PN and PO towards a low-mass protostellar outflow.  The kinematics of the phosphorus molecule lines in these two sources are quite different: the PO and PN line widths are higher in L1157-B1 than in B1-a ($\sim$6 km/s vs. 1.5 km/s), and the line wings extend to higher velocities in L1157-B1 than in B1-a (-15 km/s vs. -3 km/s).  Whether these kinematic differences correspond to different outflow energetics is unclear.  It is, however, curious that the relative kinematics of PN and PO differ between the two sources.  The PO peak is blue-shifted compared to PN in L1157-B1, which \citet{Lefloch2016} interpret as a delayed gas-phase formation of PO, while PO and PN peak at the same velocity in B1-a, indicative of joint formation or release.  With only two sources it is difficult to determine the origin of this difference, but it is suggestive that different outflow conditions can impact the desorption efficiency of different phosphorus carriers, or the efficiency of subsequent gas-phase phosphorus chemistry.

\subsection{PO/PN ratio: comparison to other observations and models}
\label{section:disc_compare}

PO has been detected towards just four other ISM regions: the L1157-B1 outflow \citep{Lefloch2016}, the massive star-forming regions W51 and W3(OH) \citep{Rivilla2016}, and the galactic center GMC G+0.693-0.03 \citep{Rivilla2018}.  In all cases the PO/PN ratio ranges from 1--3.  In B1-a, we derive a PO/PN ratio that ranges from 1--3 depending on the assumed source size.

There have been several recent models that aim to reproduce the observed PO/PN ratios of in different star-forming environments.  \citet{Jimenez-Serra2018} explore the phosphorus chemistry under a number of different energetic conditions, including heating, shocking, UV irradiation, and cosmic ray irradiation; \citet{Lefloch2016} model the shocked outflow L1157-B1; and \citet{Rivilla2016} model the collapse and warm-up chemistry in massive cores.  In all cases, the models can reproduce a PO/PN ratio of $\sim$1--3 under certain specific conditions, but also predict significant regions of parameter space in which the PO/PN ratio is much greater or much less than unity.  That we consistently observe a PO/PN ratio around unity in a variety of astrophysical sources, when a wide range of values should be possible, suggests that our understanding of the phosphorus chemistry is incomplete.  Current models consider purely gas-phase formation mechanisms for PO and PN, which results in orders of magnitude variation in the PO/PN ratio depending on the model details.  We speculate that PO and PN are more directly connected to the grain surface carriers than this, which could explain the consistency in the PO/PN ratio in a wide range of environments.  In this regard, exploring additional regimes in the phosphorus chemical network would be worthwhile; this could include an expanded ice-phase phosphorus chemical network, as well as tracking gas-phase chemistry following sputtering of other solid phosphorus carriers like larger phosphorus oxides or mineral phases of phosphorus \citep[see e.g.][]{Pasek2019}.

Currently, sources with PO non-detections have PO/PN upper limits of at least 1.3 \citep{Rivilla2016, Rivilla2018}.  More constraining PO upper limits would help to evaluate whether the PO/PN ratio is around unity in these sources, or if they represent a different chemical regime with PO/PN $<<1$.  It is also imperative to detect PO and PN towards additional sources to confirm whether the narrow range of PO/PN values persists across different environments, since our current interpretations may be biased by small-number statistics.  Additionally, since PO collisional rates are not yet available, all analyses to date have relied on LTE treatment.  Forthcoming collisional rates \citep{Lique2018} will hopefully allow for a more sophisticated derivation of PO column densities and in turn PO/PN ratios.

\subsection{Comparison to the Solar nebula}
\label{section:abund}

While there has long been evidence for heavy phosphorus depletion in the dense ISM \citep[e.g.][]{Turner1987}, B1-a is just the second low-mass star forming region with phosphorus molecule detections, and provides further insight into the specific chemical conditions of Solar-type star formation.  In B1-a, we find that the phosphorus carriers PO and PN together contain a P/H abundance of $\sim$10$^{-10}$--10$^{-9}$ (Section \ref{sec:res_cd}), corresponding to 0.05-- 0.5\% of the solar phosphorus abundance.  In the L1157-B1 outflow, also a low-mass star forming region, the P/H abundance in PO and PN is similarly on the order 10$^{-9}$  \citep{Lefloch2016}.  Thus, existing observations point to significant phosphorus depletion from the gas in low-mass protostars, implying that the majority of phosphorus is sequestered in solids in this stage.

Measurements of primitive Solar system bodies point to a similar partitioning of phosphorus in the Solar nebula as in B1-a and L1157-B1.  CI chondrites contain a nearly solar phosphorus abundance \citep{Lodders2003}, indicating a dominantly refractory reservoir of phosphorus in the young Solar system.  Indeed, volatile phosphorus was only recently detected in a comet for the first time, as part of the Rosetta mission to comet 67P/Churyumov-Gerasimenko.  PO was identified as the main phosphorus carrier in the coma, with a trace abundance of 0.011\% with respect to H$_2$O \citep{Rubin2019}.  Thus, in both low-mass protostars and the Solar nebula, it appears that the bulk of the phosphorus is contained in solids, with a very trace volatile component.  Interestingly, the carriers of volatile phosphorus may be different between protostars and comets: existing measurements of the PO/PN ratio in protostars are at most $\sim$3, while the PO/PN ratio in comet 67P has a lower limit of 10 \citep{Altwegg2019}.  Thus, chemistry may alter the importance of different volatile phosphorus carriers at different stages in the star formation sequence. 

\section{Conclusions}

We present detections of the phosphorus carriers PO and PN towards the Solar-type protostar B1-a using the IRAM 30m telescope.  A rotational diagram analysis yields column densities of $\sim$10$^{11}$--10$^{13}$ cm$^{-2}$ for PN and 10$^{12}$--10$^{13}$ cm$^{-2}$ for PO, and rotational temperatures around 6--10 K for both molecules.  Radiative transfer modeling of PN indicates that the emission is likely sub-thermal, but that the derived column densities are not severely impacted by non-LTE effects; beam dilution effects are the dominant uncertainty in deriving column densities.  The total P/H abundance in PO and PN is $\sim$10$^{-10}$--10$^{-9}$, suggesting heavy phosphorus depletion into solids.  Based on meteoritic and cometary evidence, the young Solar system and B1-a seem to have a similar partitioning of phosphorus between the refractory and volatile phases, though the volatile phosphorus carriers may differ in comets and protostars.

Comparing the IRAM 30m line profiles of PO and PN with the shock tracers SiO, SO$_2$, and CH$_3$OH, we find that the phosphorus molecules seem to originate from shocked gas, likely coincident with an outflow traced by SiO in complementary NOEMA observations.  The presence of an outflow seems to be important for phosphorus molecule release in low-mass star-forming regions, though comparing B1-a with L1157-B1, different outflow conditions may impact the observed phosphorus chemistry.  Exploring phosphorus molecule emission in a diverse sample of outflow sources would help to clarify the relationship between outflow physics and phosphorus molecule desorption and subsequent chemistry.

Like all other dense ISM sources where PO has been detected, we find a PO/PN ratio between 1 and 3.  While chemical models are able to reproduce PO/PN ratios $\sim$1 for certain conditions, they also predict significant parameter space in which PO/PN is much greater or less than unity.  That PO/PN is consistently measured within a fairly narrow range from 1--3 suggests that we have an incomplete understanding of the main phosphorus carriers in ISM grains and ices.

\section*{Acknowledgements}
This work is based on observations carried out under project numbers 006-13 and 097-17 with the IRAM 30m telescope and project number W16AN with the IRAM NOEMA interferometer. IRAM is supported by INSU/CNRS (France), MPG (Germany) and IGN (Spain).  This work was supported by an award from the Simons Foundation (SCOL \# 321183, KO).

\software{
{\fontfamily{qcr}\selectfont NumPy} \citep{VanDerWalt2011},
{\fontfamily{qcr}\selectfont Matplotlib} \citep{Hunter2007},
{\fontfamily{qcr}\selectfont Astropy} \citep{Astropy2013}, 
{\fontfamily{qcr}\selectfont emcee} \citep{Foreman-Mackey2013},
{\fontfamily{qcr}\selectfont RADEX} \citep{vanDerTak2007}
}

\FloatBarrier
\newpage

\appendix
\section{Rotational Diagram fitting}
\label{sec:app_rd}
The population of molecules in the upper state of a transition, $N_u$, is related to the rotational temperature $T_R$ and total column density $N_T$ by:

\begin{equation}
\frac{N_u}{g_u} = \frac{N_T}{Q(T_R)} e^{-E_u/T_R},
\label{eq:rd}
\end{equation}

\noindent where $g_u$ is the upper state degeneracy, $Q(T_R)$ is the molecular partition function, and $E_u$ is the upper state energy (in K).  The observed population of molecules in the upper state of a transition $N_{u,obs}$ can be found from the velocity-integrated main-beam temperature $\int T_{mb}dV$ according to:

\begin{equation}
\frac{N_{u, obs}}{g_u} = \frac{3 k_B \int T_{mb}dV}{8\pi^3\nu S\mu^2},
\label{eq:nu_obs}
\end{equation}

\noindent where $k_B$ is the Boltzmann constant, $\nu$ is the transition frequency, and $S\mu^2$ is the line intensity.  If the line is optically thick and the emission does not fill the beam, the true upper level population $N_u$ is related to $N_{u,obs}$ by:

\begin{equation}
N_{u} = N_{u,obs} \frac{C_\tau}{\eta_{bf}},
\label{eq:tau_bf}
\end{equation}

\noindent where $C_\tau$ is the optical depth correction factor $\frac{\tau}{1-e^{-\tau}}$ and $\eta_{bf}$ is the beam dilution factor (Equation \ref{eq:bf}).  The optical depth of a line $\tau$ is related to $N_u$ by:

\begin{equation}
\tau = \frac{c^3 A_{ul}N_u}{8\pi \nu^3 \Delta V}(e^{h\nu/k_B T_R} - 1),
\end{equation}

\noindent where $c$ is the speed of light, $A_{ul}$ is the Einstein coefficient, and $\Delta V$ is the line full width half-maximum.

Thus, with knowledge of the beam dilution factor, we can generate synthetic upper level populations with $N_T$ and $T_R$ as free parameters:

\begin{equation}
\frac{N_{u,obs}}{g_u} = \frac{N_T}{Q(T_R)} e^{-E_u/T_R} \frac{\eta_{bf}}{C_\tau}.
\label{eq:fit}
\end{equation}

\begin{figure*}[h]
\begin{centering}
	\includegraphics[width=0.8\linewidth]{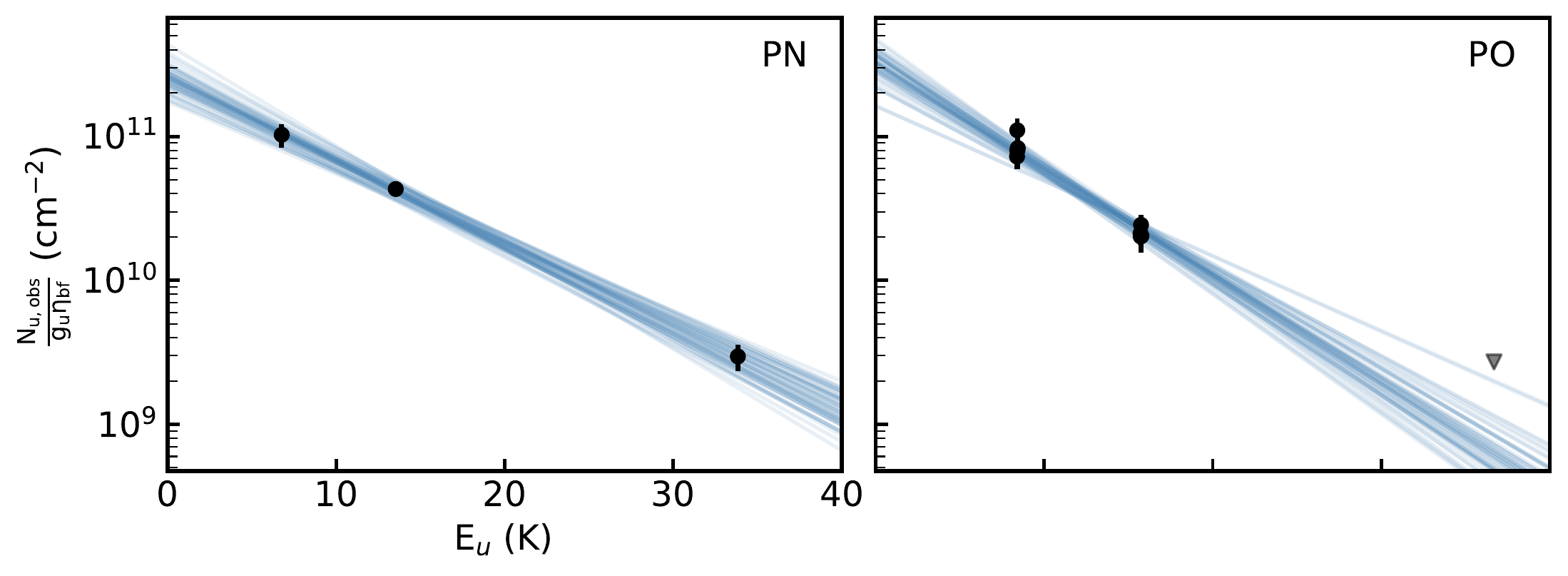}
	\caption{Example rotational diagrams for PN (left) and PO (right) assuming $R_{src}$ = 4\arcsec.  Data points are shown as black dots, and blue lines are draws from the fit posteriors.  For PO, the upper limit for the stacked 1 mm lines is shown as a grey triangle, and is not included in the fit.}
	\end{centering} 
	\label{fig:app_rd}
\end{figure*}

\noindent We fit Equation \ref{eq:fit} to the observed upper level populations (Equation \ref{eq:nu_obs}) using the MCMC package \texttt{emcee} \citep{Foreman-Mackey2013} to sample the posterior distribution.  An example rotational diagram for $R_{src}$ = 4\arcsec is shown in Figure \ref{fig:app_rd}.  We note that the PO upper limit corresponds to the stacked 1 mm lines described in Section \ref{sec:res_dets}.  The data point is not included in the rotational diagram fit, but is consistent with the fit to the 2 mm and 3 mm lines. 

\section{NOEMA observations}
\label{sec:app_noema}

The SiO images shown in Figure \ref{fig:kinematics} were taken as part of the NOEMA program W16AN (PI: T. Rice).  Full observational details can be found in T. Rice et al. (in preparation).  Briefly, Band 1 (3mm) observations containing the SiO 2--1 line were taken in the compact ``C'' configuration on 2016 December 09, with an on-source time of 12 minutes.  Band 3 (1mm) observations containing the SiO 6--5 line were taken in the sub-compact ``D'' configuration on 2017 January 23, with an on-source time of 42 minutes.  The WideX correlator was used for both lines, resulting in a spectral resolution of 6.9 km s$^{-1}$ in Band 1 and 2.3 km s$^{-1}$ in Band 3.  The GILDAS programs CLIC and MAPPING were used to produce continuum-subtracted visibilities and spectral line cubes with the CLEAN algorithm.  The restoring beam dimensions are 2.9$\times$2.1\arcsec for the SiO 2--1 line and 1.5$\times$1.1\arcsec for the SiO 6--5 line.

\section{CO and H$_2$ column densities}
\label{sec:app_CO}

\begin{deluxetable}{lcrrr}[h]
	\tablecaption{C$^{17}$O spectral line parameters$^a$ \label{tab:linedat_co}}
	\tablecolumns{5} 
	\tablewidth{\textwidth} 
	\tablehead{
		\colhead{Frequency (GHz)}       &
		\colhead{Transition}         &
		\colhead{$E_u$ (K)}              &
		\colhead{S$\mu^2$ (D$^2$)}        &
		\colhead{$g_u$}              }                                        
\startdata
112.358777 & J = 1--0, F = $\frac{3}{2}$--$\frac{5}{2}$ & 5.39 & 0.0162 & 4 \\
112.358982 & J = 1--0, F = $\frac{7}{2}$--$\frac{5}{2}$ & 5.39 & 0.0324 & 8 \\
112.360007$^b$ & J = 1--0, F = $\frac{5}{2}$--$\frac{5}{2}$ & 5.39 & 0.0243 & 6 \\
 \enddata
\tablenotetext{}{($a$) All line parameters are taken from the CDMS catalogue. ($b$) Treated as the main hyperfine component for the fitting.}
\end{deluxetable}

\begin{figure}[h]
\begin{centering}
	\includegraphics[clip, width=0.4\linewidth]{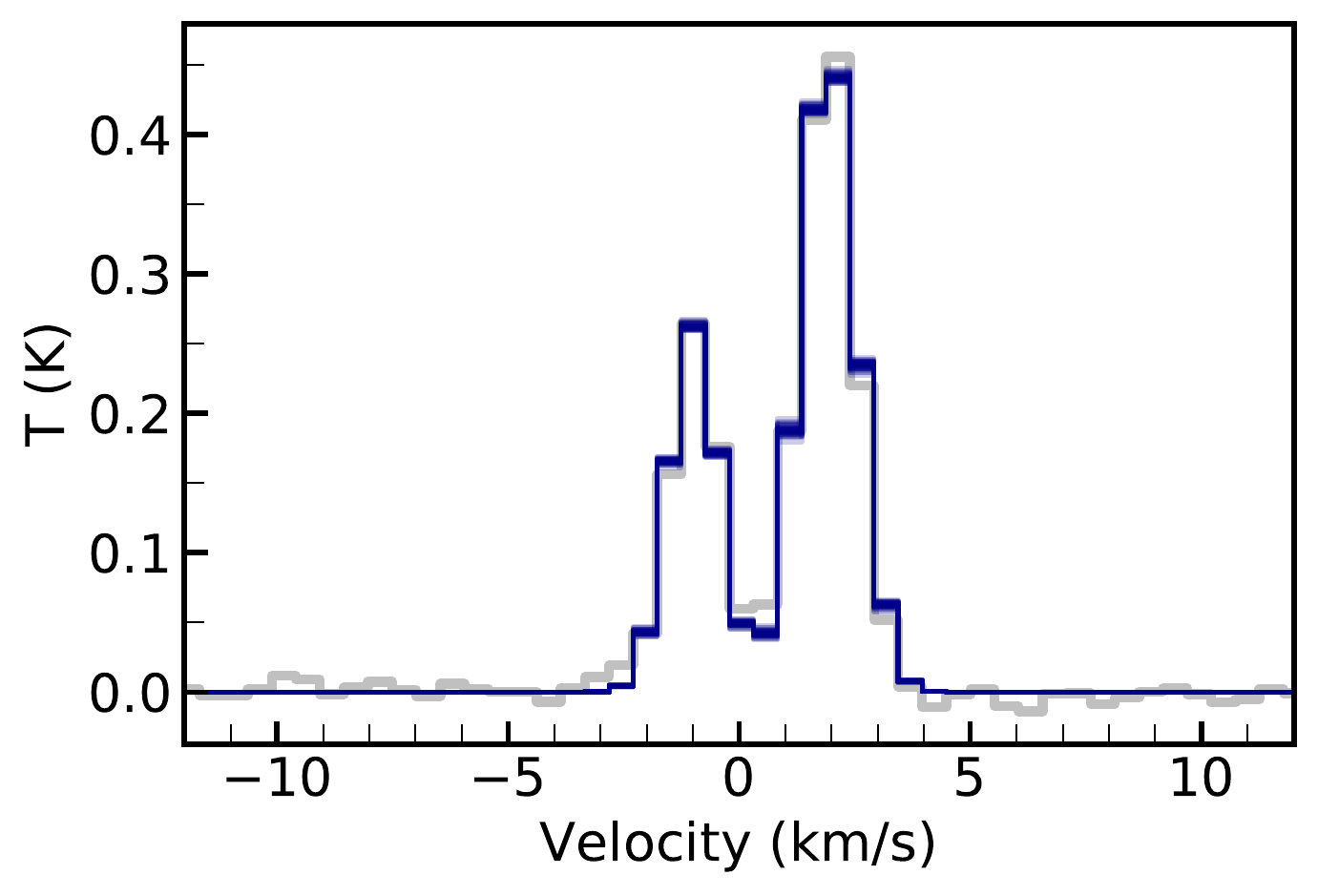}
\caption{C$^{17}$O 1--0 spectrum in B1-a, consisting of three hyperfine components at -0.98, 1.75, and 2.30 km s$^{-1}$.  The observed spectrum is shown in grey, and draws from the fit posteriors are shown in blue. }
	\end{centering} 
	\label{fig:app_co_spec}
\end{figure}

CO column densities are estimated by fitting the C$^{17}$O 1--0 hyperfine structure.  The procedure used to model the hyperfine spectrum is described in detail in \citet{Bergner2019}.  C$^{17}$O spectral line parameters are taken from the CDMS catalog using measurements from \citet{Klapper2003}, and are listed in Table \ref{tab:linedat_co}.  We use the MCMC package \texttt{emcee} \citep{Foreman-Mackey2013} to sample the posterior distribution for the fit parameters.  Figure \ref{fig:app_co_spec} shows the observed C$^{17}$O 1--0 spectrum along with draws from the fit posteriors.  

We conservatively assume a beam dilution factor of unity to derive the column density, since CO is likely present throughout the protostellar envelope; resolved observations are required for a more sophisticated treatment.  The resulting C$^{17}$O excitation temperatures and column densities (with 1$\sigma$ uncertainties) are 32 $_{-17}^{+80}$ K and 3.3 $_{-1.2}^{+6.6}$ $\times$10$^{14}$ cm$^{-2}$.  If C$^{17}$O does not fill the beam, then the true column density and excitation temperature would increase.  We note also that our fitting confirms a low optical depth ($\tau$ = 0.03 $_{-0.02}^{+0.04}$) for the C$^{17}$O J = 1--0 transition.

To convert from a C$^{17}$O column density to an H$_2$ column density, we assume the local ISM CO/C$^{17}$O ratio of 2005 \citep{Wilson1999} and a standard H$_2$/CO ratio of 10$^{4}$.  This results in an H$_2$ column density of 6.6 $_{-2.5}^{+13}$ $\times$10$^{21}$ cm$^{-2}$.

\FloatBarrier
\bibliography{references}

\begin{thebibliography}{}
\expandafter\ifx\csname natexlab\endcsname\relax\def\natexlab#1{#1}\fi
\providecommand{\url}[1]{\href{#1}{#1}}
\providecommand{\dodoi}[1]{doi:~\href{http://doi.org/#1}{\nolinkurl{#1}}}
\providecommand{\doeprint}[1]{\href{http://ascl.net/#1}{\nolinkurl{http://ascl.net/#1}}}
\providecommand{\doarXiv}[1]{\href{https://arxiv.org/abs/#1}{\nolinkurl{https://arxiv.org/abs/#1}}}

\bibitem[{{Ag{\'u}ndez} {et~al.}(2007){Ag{\'u}ndez}, {Cernicharo}, \&
  {Gu{\'e}lin}}]{Agundez2007}
{Ag{\'u}ndez}, M., {Cernicharo}, J., \& {Gu{\'e}lin}, M. 2007, \apjl, 662, L91,
  \dodoi{10.1086/519561}

\bibitem[{{Altwegg} {et~al.}(2019){Altwegg}, {Balsiger}, \&
  {Fuselier}}]{Altwegg2019}
{Altwegg}, K., {Balsiger}, H., \& {Fuselier}, S.~A. 2019, arXiv e-prints,
  arXiv:1908.04046.
\newblock \doarXiv{1908.04046}

\bibitem[{{Asplund} {et~al.}(2009){Asplund}, {Grevesse}, {Sauval}, \&
  {Scott}}]{Asplund2009}
{Asplund}, M., {Grevesse}, N., {Sauval}, A.~J., \& {Scott}, P. 2009, \araa, 47,
  481, \dodoi{10.1146/annurev.astro.46.060407.145222}

\bibitem[{{Astropy Collaboration} {et~al.}(2013){Astropy Collaboration},
  {Robitaille}, {Tollerud}, {Greenfield}, {Droettboom}, {Bray}, {Aldcroft},
  {Davis}, {Ginsburg}, {Price-Whelan}, {Kerzendorf}, {Conley}, {Crighton},
  {Barbary}, {Muna}, {Ferguson}, {Grollier}, {Parikh}, {Nair}, {Unther},
  {Deil}, {Woillez}, {Conseil}, {Kramer}, {Turner}, {Singer}, {Fox}, {Weaver},
  {Zabalza}, {Edwards}, {Azalee Bostroem}, {Burke}, {Casey}, {Crawford},
  {Dencheva}, {Ely}, {Jenness}, {Labrie}, {Lim}, {Pierfederici}, {Pontzen},
  {Ptak}, {Refsdal}, {Servillat}, \& {Streicher}}]{Astropy2013}
{Astropy Collaboration}, {Robitaille}, T.~P., {Tollerud}, E.~J., {et~al.} 2013,
  \aap, 558, A33, \dodoi{10.1051/0004-6361/201322068}

\bibitem[{{Bailleux} {et~al.}(2002){Bailleux}, {Bogey}, {Demuynck}, {Liu}, \&
  {Walters}}]{Bailleux2002}
{Bailleux}, S., {Bogey}, M., {Demuynck}, C., {Liu}, Y., \& {Walters}, A. 2002,
  Journal of Molecular Spectroscopy, 216, 465, \dodoi{10.1006/jmsp.2002.8665}

\bibitem[{{Bergner} {et~al.}(2019){Bergner}, {{\"O}berg}, {Bergin}, {Loomis},
  {Pegues}, \& {Qi}}]{Bergner2019}
{Bergner}, J.~B., {{\"O}berg}, K.~I., {Bergin}, E.~A., {et~al.} 2019, The
  Astrophysical Journal, 876, 25, \dodoi{10.3847/1538-4357/ab141e}

\bibitem[{{Cazzoli} {et~al.}(2006){Cazzoli}, {Cludi}, \&
  {Puzzarini}}]{Cazzoli2006}
{Cazzoli}, G., {Cludi}, L., \& {Puzzarini}, C. 2006, Journal of Molecular
  Structure, 780, 260, \dodoi{10.1016/j.molstruc.2005.07.010}

\bibitem[{{Chandler} \& {Richer}(2000)}]{Chandler2000}
{Chandler}, C.~J., \& {Richer}, J.~S. 2000, \apj, 530, 851,
  \dodoi{10.1086/308401}

\bibitem[{{Dubernet} {et~al.}(2013){Dubernet}, {Alexander}, {Ba},
  {Balakrishnan}, {Balan{\c{c}}a}, {Ceccarelli}, {Cernicharo}, {Daniel},
  {Dayou}, {Doronin}, {Dumouchel}, {Faure}, {Feautrier}, {Flower}, {Grosjean},
  {Halvick}, {K{\l}os}, {Lique}, {McBane}, {Marinakis}, {Moreau}, {Moszynski},
  {Neufeld}, {Roueff}, {Schilke}, {Spielfiedel}, {Stancil}, {Stoecklin},
  {Tennyson}, {Yang}, {Vasserot}, \& {Wiesenfeld}}]{Dubernet2013}
{Dubernet}, M.~L., {Alexander}, M.~H., {Ba}, Y.~A., {et~al.} 2013, \aap, 553,
  A50, \dodoi{10.1051/0004-6361/201220630}

\bibitem[{{Fontani} {et~al.}(2016){Fontani}, {Rivilla}, {Caselli}, {Vasyunin},
  \& {Palau}}]{Fontani2016}
{Fontani}, F., {Rivilla}, V.~M., {Caselli}, P., {Vasyunin}, A., \& {Palau}, A.
  2016, \apjl, 822, L30, \dodoi{10.3847/2041-8205/822/2/L30}

\bibitem[{{Foreman-Mackey} {et~al.}(2013){Foreman-Mackey}, {Hogg}, {Lang}, \&
  {Goodman}}]{Foreman-Mackey2013}
{Foreman-Mackey}, D., {Hogg}, D.~W., {Lang}, D., \& {Goodman}, J. 2013, \pasp,
  125, 306, \dodoi{10.1086/670067}

\bibitem[{{Goldsmith} \& {Langer}(1999)}]{Goldsmith1999}
{Goldsmith}, P.~F., \& {Langer}, W.~D. 1999, \apj, 517, 209,
  \dodoi{10.1086/307195}

\bibitem[{{Graninger} {et~al.}(2016){Graninger}, {Wilkins}, \&
  {{\"O}berg}}]{Graninger2016}
{Graninger}, D.~M., {Wilkins}, O.~H., \& {{\"O}berg}, K.~I. 2016, \apj, 819,
  140, \dodoi{10.3847/0004-637X/819/2/140}

\bibitem[{{Hatchell} {et~al.}(2007){Hatchell}, {Fuller}, {Richer}, {Harries},
  \& {Ladd}}]{Hatchell2007}
{Hatchell}, J., {Fuller}, G.~A., {Richer}, J.~S., {Harries}, T.~J., \& {Ladd},
  E.~F. 2007, \aap, 468, 1009, \dodoi{10.1051/0004-6361:20066466}

\bibitem[{{Hunter}(2007)}]{Hunter2007}
{Hunter}, J.~D. 2007, Computing in Science and Engineering, 9, 90,
  \dodoi{10.1109/MCSE.2007.55}

\bibitem[{{Jim{\'e}nez-Serra} {et~al.}(2018){Jim{\'e}nez-Serra}, {Viti},
  {Qu{\'e}nard}, \& {Holdship}}]{Jimenez-Serra2018}
{Jim{\'e}nez-Serra}, I., {Viti}, S., {Qu{\'e}nard}, D., \& {Holdship}, J. 2018,
  \apj, 862, 128, \dodoi{10.3847/1538-4357/aacdf2}

\bibitem[{{J{\o}rgensen} {et~al.}(2002){J{\o}rgensen}, {Sch{\"o}ier}, \& {van
  Dishoeck}}]{Jorgensen2002}
{J{\o}rgensen}, J.~K., {Sch{\"o}ier}, F.~L., \& {van Dishoeck}, E.~F. 2002,
  \aap, 389, 908, \dodoi{10.1051/0004-6361:20020681}

\bibitem[{{Kawaguchi} {et~al.}(1983){Kawaguchi}, {Saito}, \&
  {Hirota}}]{Kawaguchi1983}
{Kawaguchi}, K., {Saito}, S., \& {Hirota}, E. 1983, \jcp, 79, 629,
  \dodoi{10.1063/1.445810}

\bibitem[{{Klapper} {et~al.}(2003){Klapper}, {Surin}, {Lewen}, {M{\"u}ller},
  {Pak}, \& {Winnewisser}}]{Klapper2003}
{Klapper}, G., {Surin}, L., {Lewen}, F., {et~al.} 2003, \apj, 582, 262,
  \dodoi{10.1086/344615}

\bibitem[{{Lefloch} {et~al.}(2016){Lefloch}, {Vastel}, {Viti}, {Jimenez-Serra},
  {Codella}, {Podio}, {Ceccarelli}, {Mendoza}, {Lepine}, \&
  {Bachiller}}]{Lefloch2016}
{Lefloch}, B., {Vastel}, C., {Viti}, S., {et~al.} 2016, \mnras, 462, 3937,
  \dodoi{10.1093/mnras/stw1918}

\bibitem[{{Lique} {et~al.}(2018){Lique}, {Jim{\'e}nez-Serra}, {Viti}, \&
  {Marinakis}}]{Lique2018}
{Lique}, F., {Jim{\'e}nez-Serra}, I., {Viti}, S., \& {Marinakis}, S. 2018,
  Physical Chemistry Chemical Physics (Incorporating Faraday Transactions), 20,
  5407, \dodoi{10.1039/C7CP05605B}

\bibitem[{{Lodders}(2003)}]{Lodders2003}
{Lodders}, K. 2003, The Astrophysical Journal, 591, 1220,
  \dodoi{10.1086/375492}

\bibitem[{{Maci{\'a}}(2005)}]{Macia2005}
{Maci{\'a}}, E. 2005, {Chem. Soc. Rev.}, 34, 691, \dodoi{10.1039/B416855K}

\bibitem[{{Matthews} {et~al.}(1987){Matthews}, {Feldman}, \&
  {Bernath}}]{Matthews1987}
{Matthews}, H.~E., {Feldman}, P.~A., \& {Bernath}, P.~F. 1987, \apj, 312, 358,
  \dodoi{10.1086/164881}

\bibitem[{{Milam} {et~al.}(2008){Milam}, {Halfen}, {Tenenbaum}, {Apponi},
  {Woolf}, \& {Ziurys}}]{Milam2008}
{Milam}, S.~N., {Halfen}, D.~T., {Tenenbaum}, E.~D., {et~al.} 2008, \apj, 684,
  618, \dodoi{10.1086/589135}

\bibitem[{{Mininni} {et~al.}(2018){Mininni}, {Fontani}, {Rivilla},
  {Beltr{\'a}n}, {Caselli}, \& {Vasyunin}}]{Mininni2018}
{Mininni}, C., {Fontani}, F., {Rivilla}, V.~M., {et~al.} 2018, \mnras, 476,
  L39, \dodoi{10.1093/mnrasl/sly026}

\bibitem[{{M{\"u}ller} {et~al.}(2005){M{\"u}ller}, {Schl{\"o}der}, {Stutzki},
  \& {Winnewisser}}]{Muller2005}
{M{\"u}ller}, H.~S.~P., {Schl{\"o}der}, F., {Stutzki}, J., \& {Winnewisser}, G.
  2005, Journal of Molecular Structure, 742, 215,
  \dodoi{10.1016/j.molstruc.2005.01.027}

\bibitem[{{M{\"u}ller} {et~al.}(2001){M{\"u}ller}, {Thorwirth}, {Roth}, \&
  {Winnewisser}}]{Muller2001}
{M{\"u}ller}, H.~S.~P., {Thorwirth}, S., {Roth}, D.~A., \& {Winnewisser}, G.
  2001, \aap, 370, L49, \dodoi{10.1051/0004-6361:20010367}

\bibitem[{{Ortiz-Le{\'o}n} {et~al.}(2018){Ortiz-Le{\'o}n}, {Loinard}, {Dzib},
  {Kounkel}, {Galli}, {Tobin}, {Evans}, {Hartmann}, {Rodr{\'{\i}}guez},
  {Brice{\~n}o}, {Torres}, \& {Mioduszewski}}]{Ortiz2018}
{Ortiz-Le{\'o}n}, G.~N., {Loinard}, L., {Dzib}, S.~A., {et~al.} 2018, \apjl,
  869, L33, \dodoi{10.3847/2041-8213/aaf6ad}

\bibitem[{{Pasek}(2019)}]{Pasek2019}
{Pasek}, M.~A. 2019, \icarus, 317, 59, \dodoi{10.1016/j.icarus.2018.07.011}

\bibitem[{{Rivilla} {et~al.}(2016){Rivilla}, {Fontani}, {Beltr{\'a}n},
  {Vasyunin}, {Caselli}, {Mart{\'\i}n-Pintado}, \& {Cesaroni}}]{Rivilla2016}
{Rivilla}, V.~M., {Fontani}, F., {Beltr{\'a}n}, M.~T., {et~al.} 2016, \apj,
  826, 161, \dodoi{10.3847/0004-637X/826/2/161}

\bibitem[{{Rivilla} {et~al.}(2018){Rivilla}, {Jim{\'e}nez-Serra}, {Zeng},
  {Mart{\'\i}n}, {Mart{\'\i}n-Pintado}, {Armijos-Abenda{\~n}o}, {Viti},
  {Aladro}, {Riquelme}, \& {Requena-Torres}}]{Rivilla2018}
{Rivilla}, V.~M., {Jim{\'e}nez-Serra}, I., {Zeng}, S., {et~al.} 2018, \mnras,
  475, L30, \dodoi{10.1093/mnrasl/slx208}

\bibitem[{{Rubin} {et~al.}(2019){Rubin}, {Altwegg}, {Balsiger}, {Berthelier},
  {Combi}, {De Keyser}, {Drozdovskaya}, {Fiethe}, {Fuselier}, {Gasc},
  {Gombosi}, {H{\"a}nni}, {Hansen}, {Mall}, {R{\`e}me}, {Schroeder},
  {Schuhmann}, {S{\'e}mon}, {Waite}, {Wampfler}, \& {Wurz}}]{Rubin2019}
{Rubin}, M., {Altwegg}, K., {Balsiger}, H., {et~al.} 2019, Monthly Notices of
  the Royal Astronomical Society, 2020, \dodoi{10.1093/mnras/stz2086}

\bibitem[{{Tenenbaum} {et~al.}(2007){Tenenbaum}, {Woolf}, \&
  {Ziurys}}]{Tenenbaum2007}
{Tenenbaum}, E.~D., {Woolf}, N.~J., \& {Ziurys}, L.~M. 2007, \apjl, 666, L29,
  \dodoi{10.1086/521361}

\bibitem[{{Tobo{\l}a} {et~al.}(2007){Tobo{\l}a}, {K{\l}os}, {Lique},
  {Cha{\l}asi{\'n}ski}, \& {Alexander}}]{Tobola2007}
{Tobo{\l}a}, R., {K{\l}os}, J., {Lique}, F., {Cha{\l}asi{\'n}ski}, G., \&
  {Alexander}, M.~H. 2007, \aap, 468, 1123, \dodoi{10.1051/0004-6361:20077339}

\bibitem[{{Turner} \& {Bally}(1987)}]{Turner1987}
{Turner}, B.~E., \& {Bally}, J. 1987, \apjl, 321, L75, \dodoi{10.1086/185009}

\bibitem[{{van der Tak} {et~al.}(2007){van der Tak}, {Black}, {Sch{\"o}ier},
  {Jansen}, \& {van Dishoeck}}]{vanDerTak2007}
{van der Tak}, F.~F.~S., {Black}, J.~H., {Sch{\"o}ier}, F.~L., {Jansen}, D.~J.,
  \& {van Dishoeck}, E.~F. 2007, \aap, 468, 627,
  \dodoi{10.1051/0004-6361:20066820}

\bibitem[{{van der Walt} {et~al.}(2011){van der Walt}, {Colbert}, \&
  {Varoquaux}}]{VanDerWalt2011}
{van der Walt}, S., {Colbert}, S.~C., \& {Varoquaux}, G. 2011, Computing in
  Science and Engineering, 13, 22, \dodoi{10.1109/MCSE.2011.37}

\bibitem[{{Wilson}(1999)}]{Wilson1999}
{Wilson}, T.~L. 1999, Reports on Progress in Physics, 62, 143,
  \dodoi{10.1088/0034-4885/62/2/002}

\bibitem[{{Yamaguchi} {et~al.}(2011){Yamaguchi}, {Takano}, {Sakai}, {Sakai},
  {Liu}, {Su}, {Hirano}, {Takakuwa}, {Aikawa}, \& {Nomura}}]{Yamaguchi2011}
{Yamaguchi}, T., {Takano}, S., {Sakai}, N., {et~al.} 2011, \pasj, 63, L37,
  \dodoi{10.1093/pasj/63.5.L37}

\bibitem[{{Ziurys} {et~al.}(2007){Ziurys}, {Milam}, {Apponi}, \&
  {Woolf}}]{Ziurys2007}
{Ziurys}, L.~M., {Milam}, S.~N., {Apponi}, A.~J., \& {Woolf}, N.~J. 2007, \nat,
  447, 1094, \dodoi{10.1038/nature05905}

\end{thebibliography}
\bibliographystyle{aasjournal}
\end{document}